# On The Impact of Merge Request Deviations on Code Review Practices

Samah Kansab
Software and IT Engineering Department
École de technologie superieure (ÉTS)
Montreal, Canada
samah.kansab.1@etsmtl.net

Francis Bordeleau
Software and IT Engineering Department
École de technologie supérieure (ÉTS)
Montreal, Canada
francis.bordeleau@etsmtl.ca

Ali Tizghadam
Network Softwarisation and AI
TELUS
Toronto, Canada
ali.tizghadam@telus.com

*Abstract*—Code review is a fundamental practice in software engineering, ensuring code quality, fostering collaboration, and reducing defects. While research has extensively examined various aspects of this process, most studies assume that all code reviews follow a standardized evaluation workflow. However, our industrial partner, which uses Merge Requests (MRs) mechanism for code review, reports that this assumption does not always hold in practice. Many MRs serve alternative purposes beyond rigorous code evaluation. These MRs often bypass the standard review process, requiring minimal oversight. We refer to these cases as deviations, as they disrupt expected workflow patterns. For example, work-in-progress (WIP) MRs may be used as draft implementations without the intention of being reviewed, MRs with huge changes are often created for code rebase, and library updates typically involve dependency version changes that require minimal or no review effort. We hypothesize that overlooking MR deviations can lead to biased analytics and reduced reliability of machine learning (ML) models used to explain the code review process.

This study addresses these challenges by first identifying MR deviations. Our findings show that deviations occur in up to 37.02% of MRs across seven distinct categories. In addition, we develop a detection approach leveraging few-shot learning, achieving up to 91% accuracy in identifying these deviations. Furthermore, we examine the impact of removing MR deviations on ML models predicting code review completion time. Removing deviations significantly enhances model performance in 53.33% of cases, with improvements of up to 2.25 times. Additionally, their exclusion significantly impacts model interpretation, strongly altering overall feature importance rankings in 47% of cases and top-k rankings in 60%.

Our contributions include: (1) a clear definition and categorization of MR deviations, (2) a novel AI-based detection method leveraging few-shot learning, and (3) an empirical analysis of their exclusion impact on ML models explaining code review completion time. Our approach helps practitioners streamline review workflows, allocate reviewer effort more effectively, and ensure more reliable insights from MR analytics.

## I. INTRODUCTION

Code review is a critical practice in industrial software development, ensuring code quality, consistency, and collaboration [43]. This process enhances code readability [3] and reduces the likelihood of introducing defects [28, 3]. Various code review mechanisms exist, such as Guerrit, Pull Requests (PRs) and Merge Requests (MRs), which facilitate the evaluation of code changes before integration. In this study, we adopt the MR mechanism, as it is the primary mechanism used by our industrial partners.

Code review has been extensively studied in the literature, with a focus on explaining, analyzing, and improving the process. For instance, one key research direction explores the time required to complete code reviews, as it reflects the effort invested in reviewing code changes [9, 14, 42, 26, 15]. Additionally, researchers have examined the time taken to receive the first response [14] and investigated the intentions behind code changes [45]. Various factors influencing review time and effort have also been explored, including tools for prioritizing code reviews [12, 35] and early prediction of whether code reviews will be merged or rejected [15].

When analyzing code reviews, researchers typically preprocess data using various methods, such as removing outliers [9], conducting correlation and redundancy analyses [27], and data annotation[46]. However, they generally assume that all collected data represents standard code review instances following a structured evaluation workflow. In practice, our industrial partners report that a significant portion of MRs deviate from this assumption. These MRs serve purposes beyond the intent of having a rigorous code evaluation, such as code cleaning, work in progress or draft, and often bypass standard review procedures with minimal oversight. We refer to these cases as **deviations** because they do not conform to the expected review workflow, and can unnecessarily divert the reviewers' attention. In this study, *we hypothesize that ignoring these deviations can introduce bias into analyses, as these data points do not accurately reflect the formal review process. This can lead to inefficiencies in assessing reviewer effort and produce misleading conclusions about the code review process, particularly in large-scale industrial settings*.

While deviations may seem similar to outliers and extremes, they represent a distinct challenge. Outliers are data points that significantly differ from typical trends of a metric, such as MRs with unusually long review times, and are often detected using techniques like Local Outlier Factor (LOF) [8], Isolation Forest [7], and One-Class SVM [29]. Extremes, on the other hand, refer to the smallest or largest values within a metric, such as particularly small or large MRs, and are identified using methods like threshold-based [6] or percentile-based

[32]. Unlike these statistical anomalies, deviations do not necessarily manifest as numerical irregularities, but as cases where MRs break expected workflow patterns *(i.e., submission of a code change for review, assignment to a reviewer, iterative feedback and revisions, final approval, and merge into the main branch)*. Additionally, some regular MRs also lack review activity due to factors such as in-person oral reviews when developers work on-site, making a purely statistical filter ineffective. Identifying deviations requires contextual analysis across multiple dimensions, including MR intent, author behaviour, and review engagement. Standard anomaly detection techniques fail to capture these nuances, making it necessary to develop a dedicated approach for recognizing deviations.

This study aims to improve the efficiency and accuracy of MR-based code review by identifying, categorizing, and automating the detection of deviations. To achieve this, we analyze a dataset of 26.7k MRs from four teams within our industrial partner—a networking software company specializing in 5G and cloud edge fabric technologies for telecom, data center, and cloud service providers. The company follows a DevOps approach to streamline software development and deployment. Our investigation focuses on three key research questions (RQs):

*RQ1: What are the common types of deviations in the MR review process?* Understanding deviations is crucial for practitioners to better categorize MRs and assess how the review process is used in practice. Our analysis shows that up to 37.02% of MRs in industrial projects are deviations, primarily consisting of library updates, configuration changes, and code cleaning tasks. Recognizing these patterns can help teams streamline the review workflow, prioritize critical MRs, and allocate reviewer effort more effectively.

*RQ2: To what extent can AI automate the detection of deviation cases in MRs?* Manually identifying deviations in large-scale projects is infeasible, as it requires domain expertise and substantial reviewer effort. Traditional rule-based methods struggle to capture the complex patterns in MR devi- ations, making AI-driven automation essential for scalability and consistency. We explore few-shot learning and BERT- based models, demonstrating that Few-shot learning models can accurately detect deviations with up to 91% accuracy using only 15 examples, significantly reducing manual workload and improving classification reliability.

*RQ3: How does excluding deviations impact ML models that predict code review completion time?* Beyond automating deviation detection, we investigate the impact of deviations on machine learning models used to predict and explain the code review completion time, as a metric commonly used in literature to assess the effort. Our findings indicate that removing deviations significantly improves or does not affect model performance in 77.57% of cases, while substantially altering feature importance rankings in 47% of cases and 60% of top-k rankings. These results highlight that neglecting deviations can mislead process optimization efforts and cause teams to draw incorrect interpretation.

Our key contributions include:
**(1)** *A clear definition and identification of MR deviations;*
**(2)** *A novel AI-driven approach for automatically detecting deviations using few-shot learning;*
**(3)** *An evaluation of how deviations impact ML models predicting time to complete code review*, demonstrating their importance in improving analytical accuracy. These insights are crucial for both practitioners and researchers aiming to refine MR evaluation processes and ensure that review efforts are directed toward the most impactful code changes.
**(4)** A replication package enabling practitioners to reproduce our approach and apply it in their own review workflows [1].

The paper is structured as follows: Section 2 presents the related work. Section 3 to 5 outlines the results, detailing the motivation, approach, and findings for each RQ. Section 6 discusses our findings and the implications of our study, and finally, Section 7 concludes the paper.

## II. RELATED WORK

Code review has been extensively studied to understand and improve its efficiency, with research employing vari- ous methodologies and data preprocessing techniques. For instance, Chouchen et al. [9] predicted code review completion time on Gerrit by removing outliers and running correlation analysis. Hasan et al. [14] found that first-response time impacts pull request lifetime, though bot-first reviews behave differently, analyzing GitHub pull request data using regression with ANOVA II to analyze feature importance. Jiang, Adams, and German [17] showed that reviewer count affects review duration, using historical review data from open-source projects to establish patterns using data discretization technique on the dependent variable. Baysal et al. [4] found that both technical and non-technical factors influence review time by combining developer surveys with repository mining. To ensure data reliability, they removed outliers from code review duration metrics, excluded instances with inactive reviewers, and focused specifically on WebCore-related code. To improve efficiency, Fan et al. [12] and Islam et al. [15] used machine learning to predict merge likelihood. Fan et al. [12] leveraged correlation analysis, redundancy analysis and feature importance techniques, while Islam et al. [15] filtered the data statistically like removing instances that are not merged. Li et al. [23] and Khatoonabadi et al. [20] addressed duplicate and abandoned pull requests to reduce wasted effort, running correlation and redundancy analysis. Wang, Bansal, and Nagappan [45] examined the intent behind code changes to predict review effort, using a heuristics-based approach to classify change intents and machine learning models trained on extracted metadata and intent features. However, their study focuses on intent categorization without considering deviations from standard review processes and their impact.

To improve dataset quality, Liu, Lin, and Thongtanunam [25] used LLMs to clean review comments, applying natural language processing for automated dataset refinement. Liang et al. [24] introduced CupCleaner, a semantic-based dataset cleaner for comment updates, enhancing dataset usability by detecting and removing outdated or redundant comments. Krutauz et al. [21] applied hierarchical clustering to reduce multicollinearity in review datasets, ensuring more reliable predictive models. Jureczko, Kajda, and Go´recki [19] analyzed comment density correlations to study knowledge sharing, using repository metadata and developer interaction logs. Gupta and Sundaresan [13] filtered non-actionable comments with regex to refine datasets for comment analysis.

---
[1] https://figshare.com/s/c00921ec9d3d091ca834

While prior studies have primarily focused on improving predictions and refining datasets, less attention has been given to understanding deviations from standard review processes. Our work addresses this gap by systematically identifying and analysing deviations in the MR review process and their impact on ML models.

## Results

To achieve our goal, we follow a rigorous methodology, as illustrated in Figure 1. Using the GitLab API, we collect data on 26.7k MRs from four teams, as summarized in Table III. This dataset enables us to address the following RQs:

### III. *RQ1: What are the common types of MR deviations in the code review process?*

#### *Motivation*

The MR mechanism is designed to facilitate code reviews, ensuring that proposed changes meet quality standards before being integrated into the codebase. However, in practice, MRs are not always used as intended. Our industrial partners have observed that, in some cases, MRs deviate from their primary purpose, serving alternative functions. These MR deviations can arise for various reasons, such as using MRs for code cleaning, work-in-progress changes, or draft development, without the expectation of review. Understanding these deviations is crucial for practitioners to better classify and assess how the MR mechanism is being used and to analyze its broader impact on the code review process. In this RQ, we aim to systematically define and identify these deviations to enable further study of their implications.

#### *Approach*

To define and identify MR deviations, we follow these steps:

#### A. Metrics Collection

To identify MR deviations, we first collect a comprehensive set of metrics that capture different aspects of the MR process, providing valuable insights into MR characteristics such as review duration, number of reviewers, and volume of comments, which may indicate deviations from standard review workflows. Table I presents each metric along with its definition and justification, explaining its relevance and ensuring that our analysis is grounded in meaningful indicators for detecting deviations.

#### B. Sample Selection

Given the impracticality of manually evaluating thousands of MRs, we adopted a representative sampling approach based on established methodologies [41, 5]. The sample size was calculated using statistical parameters, including the total number of MRs, the desired confidence level, the margin of error, and an estimated population proportion. Initially, the sample size ($s$) was determined using the formula $s = \frac{z^2 \cdot p \cdot (1-p)}{c^2}$, where $z$ is the Z-score for the specified confidence level (e.g., 1.96 for 95%), $p$ is the estimated proportion (assumed to be 0.5 for maximum variability), and $c$ is the margin of error (set to 0.05 or 5%). Given the finite MR population, a finite population correction was applied, adjusting the sample size to $ss = \frac{s \cdot P}{s+P}$, where $P$ is the total number of MRs. This correction reduces the sample size for smaller populations, ensuring accuracy under finite conditions. The final sample size was rounded up to the nearest whole number to achieve the desired statistical confidence. The sample size of each project is given in Table III.

#### C. Annotation Process

As shown in Figure 2, the annotation process for identifying and categorizing MR deviations involves three main phases: definition, annotation, and validation. These phases are iterative and collaborative, ensuring robustness and alignment with industrial insights. The process is carried out by five contributors, including three researchers and two industry partners, combining academic rigor with real-world expertise to ensure the relevance and accuracy of the categorization.

*1) Definition Phase:* The process begins by defining the practical use cases of the MR based on real-world scenarios provided by our industrial partners. Concrete MR examples are analyzed to understand how the mechanism is utilized in practice. Identified use cases are then categorized into preliminary deviation types by grouping MRs that diverge from the standard MR process into broader categories. Next, these preliminary categories are refined into formal deviation types through collaborative work sessions involving the entire team. During this phase, categories are further clarified and precisely defined to capture the nuances of each deviation. For each formal category, structured descriptions are provided to distinguish characteristics of these deviations.

*2) Annotation Phase:* The annotation phase begins with the manual labelling of randomly selected samples, as described in the previous section. Two researchers collaborate to annotate the samples, ensuring consistency with predefined deviation categories and definitions, while a third researcher supervises the process. Each MR is systematically evaluated to determine its alignment with the identified deviation types, following the formal definitions established in the earlier phases.

*3) Validation Phase:* To validate the annotation, a collaborative inspection phase is conducted, where all annotated MRs are collectively reviewed by the researchers, and random instances are investigated by industrial partners. This review addresses any ambiguities in the annotations, refines unclear definitions, and ensures consistency across all annotators.

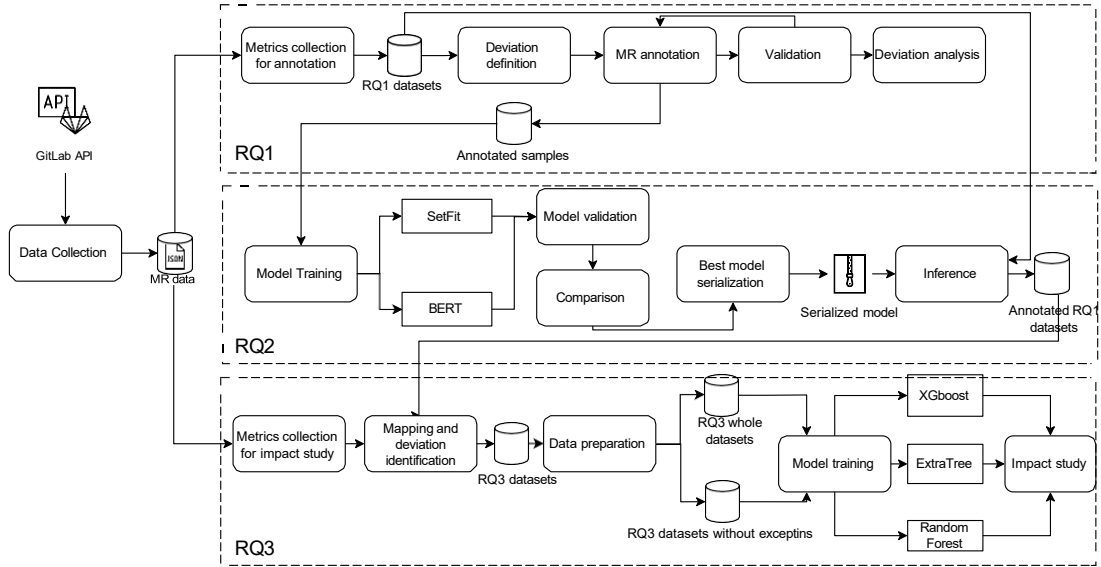

Fig. 1. Our methodology

TABLE I
METRICS FOR IDENTIFYING MR DEVIATIONS

| Metric | Definition | Justification |
| --- | --- | --- |
| Title | The title of the MR | Often hints at the MR's purpose. |
| Description | The description of the MR | Provides more context to identify MR activities. |
| File types | The list of types of MR files | Specific file types (e.g., '.md' for docs) indicate MR focus. |
| Code churn | MR total lines of code added+removed | Reflects MR complexity and review effort. |
| Additions | Lines of code added in the MR | Low or high additions may signal minor fixes or major updates. |
| Deletions | Lines of code removed in the MR | Large deletions may indicate cleanup or isolated adjustments. |
| Review duration | Time from MR creation to merge | Short reviews suggest high priority; long ones indicate low priority. |
| Reviewer participation | Number of reviewers involved | Few or no reviewers suggest the MR is low-impact or non-essential. |
| Number of commits | Total commits in the MR | High commits may indicate ongoing work or evolving updates. |
| Number of committers | Distinct contributors to the MR | Multiple committers may indicate collaborative work or task complexity. |
| Source branch | Branch from which the MR was created | Branch name may indicate experimental or feature work. |
| Target branch | Branch into which the MR will be merged | Non-main branches suggest non-critical or work-in-progress changes. |
| Commit messages | Messages associated with each commit | Can indicate purpose; non-technical tags suggest non-code MRs. |
| Number of comments | Total comments on the MR | Few comments suggest the MR was not meant for in-depth review. |
| Comment messages | Content of comments on the MR | Comments content detail the review interaction. |
| Labels | Tags or labels assigned to the MR | Labels like "documentation" or "build" indicate MR purpose. |
| Number of reviewers | Number of reviewers of the MR | Multiple reviewers indicate the MR importance |
| Reviewer comments | Content of reviewer comments | Non-technical comments suggest procedural rather than code-related focus. |
| Time to first review | Time until the first reviewer engages | Quick engagement suggests high-priority MRs. |

Throughout the process, feedback from industrial partners is incorporated to align the deviation definitions with real-world practices. This industrial input validates the practical relevance of the categories and ensures the annotations capture meaningful and actionable insights.

*4) MR Analysis:* At this stage, we systematically categorize the occurrence of MR deviations across different projects. This involves analyzing the distribution and frequency of each deviation type, assessing their prevalence within various projects.

### *Results*

**As shown in Table III, MR deviations constitute up to 37.02% of cases across seven distinct categories in the analyzed projects (Table II),** each representing a deviation from standard review practices. These deviations often indicate scenarios where a traditional, detailed code review may be de-prioritized or streamlined. By systematically categorizing these deviations, Table II provides a structured classification with clear definitions, enabling quick identification and appropriate handling of MRs that do not necessitate full code scrutiny.

**We observe in Figure 3 that library updates account for up to 39.3% of all MR deviations, particularly in the CP project,** indicating that a significant portion of MRs are created solely for routine dependency adjustments. Similarly, high proportions are observed in the MP project (39. 0%), the DP project (37. 6%) and the PF project (35. 9%). Such updates, typically involving minor version increments without impacting core functionality, do not necessitate a formal review process, as they rarely require reviewers' scrutiny. Explicitly, these MRs include 1-2 addition and 1-2 deletion

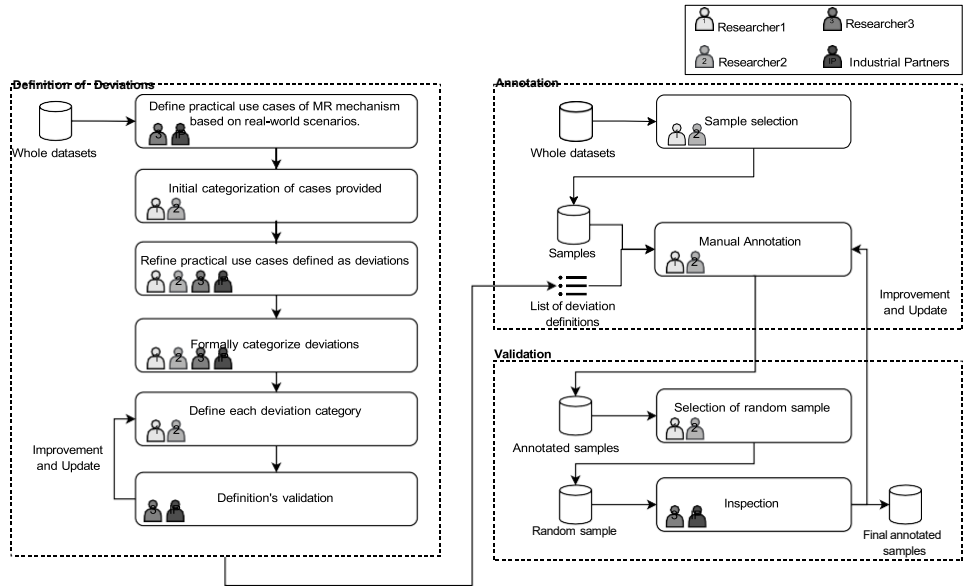

Fig. 2. The annotation process for definition, annotation and validation of MR deviations

TABLE II
DEVIATION TYPES AND DEFINITIONS

| Deviation | Definition |
| --- | --- |
| Experimental or WIP (Work in Progress) (EOW) | MRs created for experimentation or early-stage work, often marked as drafts or WIP. These MRs are not intended for immediate integration and usually do not require thorough review, as they serve as test cases, placeholders, or incomplete implementations. |
| Code Cleaning (CC) | MRs focused on removing outdated, unused, or redundant code without introducing new functionality. These changes do not require extensive review as they do not alter logic or behavior. |
| Library Update (LU)) | MRs that solely update the version of a dependency or library without modifying functionality. These changes are typically automated or routine, requiring minimal review as they do not introduce new code. |
| Build or Configuration Adjustments (BOCA) | MRs that modify CI/CD configurations, build scripts, or environment settings (e.g., Docker, Kubernetes) without impacting functional code. These changes are often minor, repetitive, and self-contained, requiring little to no review as they do not affect software behavior or introduce logical errors. |
| Revert Commits (RC) | MRs created to undo previous changes, restoring the codebase to an earlier state. These changes are typically automated and do not require extensive review, as they simply roll back prior modifications without introducing new logic. |
| Huge Changes (HC) | MRs involving large-scale modifications, such as re-branching, code restructuring, or bulk file movements. These changes are often too extensive for a detailed line-by-line review, making traditional review workflows ineffective (More than 500 commits or 10,000+ lines changed.). |
| Empty Change Set (ECS) | MRs that contain no actual modifications, often created unintentionally or due to misconfigured automation. These MRs do not require review, as they introduce no changes to the codebase. |

in 64%-86% of the cases, and no review activity (comments, assignment, or approval) for 94% -96% of the cases. This procedural use of MRs for non-critical changes dilutes the focus of the review process, introducing bias in the related analysis.

**The frequency of build or configuration adjustments and code cleaning MR deviations across projects suggests that the MR process is used for changes that do not fundamentally need formal code review.** Build or configuration adjustments deviations are particularly prevalent in the FPGA project, where they account for 35.6% of all MR deviations. Significant proportions are also observed in the DP project (22.0%), CP project (21.3%), and PF project (19.8%). The high rate of these deviations underscores how the MR process is regularly used for configuration tweaks and CI/CD adjustments, which are typically infrastructure-related and involve minimal impact on the codebase's functionality. These changes, primarily technical setup adjustments, could often bypass the MR process without compromising code quality, freeing up review resources for more critical modifications. This observation is proved by statistics showing that these MRs do not benefit from review activity for 98%-100% of the cases.

Similarly, code cleaning MR deviations—reaching up to 31.7% in the MP project—reflect an emphasis on codebase maintenance through the removal of obsolete code. Other projects also exhibit high percentages, with PF (24.4%), DP (19.3%), and FPGA (15.6%). In these MRs, we observe that the deletions dominant the additions in the churn, as no new code is added. While maintaining a clean codebase is essential, these routine tasks do not require the scrutiny of a formal review while it does not impact the code in use.

TABLE III
PROJECT TEAMS OF OUR INDUSTRIAL PARTNER

| Group | Description | #MR | #Projects | Sample size | %MR deviations |
|---|---|---|---|---|---|
| Management Plane (MP) | Manage the communication with Fabric | 6,344k | 20 | 363 | 22.59 |
| Control Plane (CP) | Orchestrate paths for packets and frames | 8,396k | 31 | 360 | 24.44 |
| Data Plane (DP) | Forward packet and frames between interfaces | 7,416k | 16 | 359 | 30.36 |
| FPGA | designs reprogrammable integrated circuits | 735 | 19 | 246 | 36.59 |
| Platform (PF) | Manage low-level platforms | 4,004k | 30 | 348 | 37.02 |

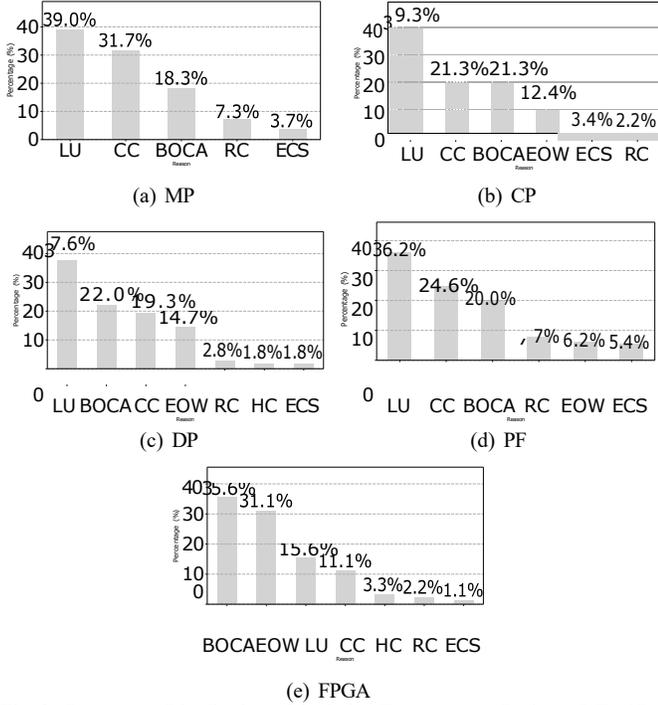

Fig. 3. Percentage Distribution of deviation Types Across Projects (MP, CP, DP, PF, FPGA).
Note: BOCA = Build or Configuration Adjustments, EOW = Experimental or Work in Progress, CC = Code Cleaning, LU = Library Update, RC = Revert Commits, HC = Huge Changes, ECS = Empty Change Set.

**Furthermore, experimental or work-in-progress MR deviations, particularly prominent in the FPGA project at 31.1%, represent another instance where MRs are often created without a genuine need for review.** These MRs, intended for preliminary testing or drafts, serve primarily as placeholders and are not meant for integration. Similar trends are observed in the CP project (12.4%) and the PF project (6.9%), indicating that MRs are frequently used to manage exploratory changes that could be more efficiently handled in isolated test branches or experimental environments.

**In contrast, MR deviations like revert commits, huge changes, quick completion, and empty change sets occur at lower rates (below 7.6% across projects)**, indicating that these are relatively rare and unlikely to disrupt the review process. The highest occurrence of revert commits deviations is seen in MP (7.3%) and PF (7.7%), while huge changes (HC) appear primarily in FPGA (3.3%) and DP (1.8%). Empty change sets (ECS) remain minimal across projects, with the highest rate at 5.3

## IV. RQ2: To what extent can AI automate the detection of deviation cases in MRs?

### Motivation

Currently, identifying deviations is time-consuming and subject to human error. By leveraging AI, we could streamline this process, enabling teams to quickly and accurately flag MRs that deviate from standard review practices and may not require full scrutiny. This automation would free up valuable reviewer time, reduce bottlenecks in the review pipeline, and ensure that attention is focused on impactful, high-priority changes. Investigating the potential for AI-driven deviation detection offers a pathway to more efficient, scalable review workflows in development teams.

### Approach

To automate MR deviation detection, we compare two AI-based approaches: BERT and SetFit. BERT is known for its high accuracy but requires extensive labeled data and computational resources. In contrast, SetFit enables efficient classification with fewer annotations, making it a suitable choice for scenarios with limited labeled data.

### A. Model Training

*1) BERT transformers:* Bidirectional Encoder Representations from Transformers (BERT) is a transformer-based model that learns contextual word representations through self-supervised pretraining on large text corpora [11]. Following previous studies [47, 37, 48], we fine-tune BERT for text classification to predict the reason for each MR based on its description.

The dataset is preprocessed by encoding target labels, tokenizing text samples using BertTokenizer, incorporating special tokens, attention masks, and padding for uniform input length. The tokenized data is processed in batches for efficient training. The model is fine-tuned using cross-entropy loss and the AdamW optimizer over multiple epochs.

*2) SetFit:* SetFit is a few-shot learning framework designed for text classification tasks where labeled data is limited [44]. In our experiment, instead of training a traditional classifier, we fine-tune a T5 model to generate the deviation (or not) for each MR. Similar to recent works [30, 36], our approach utilizes a small labeled dataset and a few training epochs(5) to minimize computational costs. The model processes input text sequences, learns meaningful representations, and generates predictions in a few-shot setting. This method enables efficient classification while reducing annotation and training overhead, making it practical for large-scale industrial datasets.

## B. Model evaluation

To evaluate model performance, we use commonly adopted classification metrics: Accuracy: Percentage of correct predictions out of total predictions. Recall: Ratio of correctly identified positive cases to actual positive cases. Precision : Proportion of true positive predictions out of all positive predictions. F1-score : Harmonic mean of precision and recall.

## C. Model validation and Comparison

To ensure reliability and account for variability, we perform 10 bootstrap iterations for each model, given the computational cost of training. We report the median results. For each iteration, we split the data into training (80%) and validation (20%) sets. To rank model performance, we use the Scott-Knott Effect Size Difference (ESD) Test[2], a hierarchical clustering method that prioritizes significant performance differences, widely used in prior studies [31, 32, 38]. The best model is then serialized to be used for the next RQ.

## Results

**As shown in Table IV, our few-shot learning approach with SetFit achieves up to 91% accuracy, outperforming BERT fine-tuning in most cases when sufficient training instances are available.** Notably, when using 15 instances per class, SetFit consistently achieves the highest performance across all projects. BERT fine-tuned for 5 epochs demonstrates strong performance in certain cases, achieving an accuracy up tp 0.84, but still does not surpass SetFit's top results. When BERT is trained for only 3 epochs, its accuracy drops significantly, down to 0.53 showing the lowest results on PF dataset, indicating that larger fine-tuning is necessary even using all the datasets for better performance.

In addition to accuracy, we observe that SetFit at 15 instances per class achieves an average F1-score of 0.84 across projects, whereas BERT-5 reaches 0.74. The precision scores for SetFit (0.75 on average) also surpass those of BERT (0.65). More notably, recall values for SetFit (0.91 at 15 instances) demonstrate superior generalization compared to BERT (0.76 at 5 epochs).

Overall, SetFit with 15 instances per class and the same number of epochs as BERT ranks highest, showing its ability to generalize well with a small labeled dataset compared to BERT. The performance gap between SetFit and BERT increases as the number of training instances grows, reaffirming SetFit's robustness in few-shot learning scenarios.

## V. RQ3: What is the Impact of Excluding MR Deviations on Model Performance and Interpretation?

**Motivation**

Code review completion time is the time elapsed between its first submission for review and its final decision. It has been extensively studied in the literature, as it serves as a key metric for assessing the effort invested in the review process. Researchers employ various data preprocessing techniques, such as outlier removal and correlation analysis, and utilize ML models to predict and explain this metric. However, deviations in the data can significantly impact the reliability of these models performance and interpretation. Such deviations can ultimately impact code review related decisions, particularly when ML models are used to explain process dynamics, potentially leading to unintended or misleading conclusions. In this RQ, we aim to quantify the impact of removing these deviations on both model performance and interpretation, providing an understanding of how these preprocessing choices shape the analysis and insights derived from ML-based code review studies.

[2]https://cran.r-project.org/web/packages/ScottKnottESD/

TABLE IV
THE MEDIAN PERFROMANCES OF EACH ALGORITHM

| Project | Instances | SetFit (5 Epochs) | | | |
|---|---|---|---|---|---|
| | | Accuracy | Precision | Recall | F1-score |
| PF | 5 | 0.600 | 0.550 | 0.650 | 0.600 |
| DP | | 0.590 | 0.520 | 0.620 | 0.570 |
| FPGA | | 0.500 | 0.450 | 0.550 | 0.500 |
| MP | | 0.420 | 0.400 | 0.500 | 0.440 |
| CP | | 0.500 | 0.450 | 0.550 | 0.500 |
| PF | 10 | 0.800 | 0.750 | 0.850 | 0.800 |
| DP | | 0.770 | 0.700 | 0.800 | 0.750 |
| FPGA | | 0.769 | 0.720 | 0.820 | 0.770 |
| MP | | 0.751 | 0.680 | 0.780 | 0.730 |
| CP | | 0.700 | 0.650 | 0.750 | 0.700 |
| PF | 15 | **0.910** | 0.780 | 0.980 | 0.870 |
| DP | | 0.870 | 0.740 | 0.940 | 0.830 |
| FPGA | | 0.800 | 0.780 | 0.980 | 0.870 |
| MP | | 0.790 | 0.710 | 0.910 | 0.800 |
| CP | | 0.900 | 0.730 | 0.930 | 0.820 |
| Project | Epochs | BERT | | | |
| | | Accuracy | Precision | Recall | F1-score |
| PF | 3 | 0.530 | 0.280 | 0.530 | 0.360 |
| DP | | 0.710 | 0.600 | 0.710 | 0.610 |
| FPGA | | 0.660 | 0.435 | 0.660 | 0.520 |
| MP | | 0.670 | 0.450 | 0.670 | 0.540 |
| CP | | 0.640 | 0.410 | 0.640 | 0.500 |
| PF | 5 | 0.810 | 0.780 | 0.810 | 0.800 |
| DP | | 0.840 | 0.820 | 0.840 | 0.800 |
| FPGA | | 0.680 | 0.610 | 0.680 | 0.630 |
| MP | | 0.790 | 0.710 | 0.790 | 0.740 |
| CP | | 0.760 | 0.670 | 0.760 | 0.700 |

**Approach**

To assess the impact of removing deviations on the models predicting code review completion, we compare models trained on the full set of MRs with those trained on a subset excluding deviations. The comparison follows these steps:

## A. Data Preparation

*1) Data Annotation:* Using the best-performing model from RQ2—SetFit with 15 instances per class—we classify the entire dataset to determine whether each MR exhibits deviation. To ensure reliability, we randomly sample and validate predictions with our industrial partners of the new predictions.

*2) Metrics Collection:* While the initial metrics were designed to predict deviations, this step focuses on extracting features that explain the code review completion time. The selected metrics, derived from prior studies [9, 14, 27, 42]. We ensured that all the metrics can be collected at the moment of creation of the MR. The metrics, with their description, are summarized in Table V.

*3) Collinearity Analysis:* To enhance model reliability and interpretability, we mitigate collinearity through correlation and redundancy analysis, following Tantithamthavorn et al. [39] and Cito et al. [10], as well as previous studies [18, 38]. First, we apply Spearman correlation analysis to remove highly correlated features, defining two metrics as correlated if their coefficient exceeds 0.70, as suggested by McIntosh et al. [28] and Lee et al. [22]. Since correlation alone does not fully eliminate collinearity, we also perform redundancy analysis using the $R^2$ value to identify independent features that can be explained by others. A threshold of 0.90 is applied, following prior work [40, 22], to filter out redundant features.

### B. Model Training and Evaluation

Following Chouchen et al. [9], we predict code review completion time, a key metric balancing prediction and interpretability. We employ ExtraTrees, XGBoost, and Random Forest, identified as top-performing algorithms in prior studies. Model performance is evaluated using SA (Standard Accuracy), MAE (Mean Absolute Error), and MSE (Mean Squared Error). For statistical robustness, we apply a 100 out-of-sample bootstrap technique [34, 39] to ensure stable conclusions. We use feature importance calculation for each of the 100 bootstrap samples.

### C. Impact Assessment

*1) Performance:* To assess the impact of removing MR deviations, we compute the ratio of the median bootstrap performance of model without deviations to that with deviations [32]. If lower values indicate better performance, we use 1 − median. For instance, if the median MSE without deviations is 0.2 and with deviations is 0.5, the improvement factor is `(1−0.2)/(1−0.5) = 1.6`. Conversely, if SA improves from 0.4 to 0.8, the improvement is `0.8/0.4 = 2` times. For significance testing, we apply Cliff's Delta and Wilcoxon tests between the 100 bootstrap results for each model.

*2) Interpretation:* For each model (those with and those without deviations), we use feature importance calculation for each of the 100 bootstrap samples, we then compute rankings per bootstrap and aggregate them using Scott-Knott ESD to derive a single rank per model. To assess rank similarity, we apply Kendall's Tau (*τ*), previously used in [33, 2, 1]. This method uses Spearman's rank correlation to compare the similarity between two feature importance ranks. A Tau correlation of 1 indicates total similarity, while -1 indicates total dissimilarity. For the top-5 features, we use the Top-K overlap method [16], as used by Rajbahadur et al. [33], measuring intersection divided by union for top-1, top-3, and top-5 features across iterations.

### *Results*

**Removing deviations results in significant improvements in 53.33% of cases, with a large effect size and a p-value<0.05, indicating statistical significance. In 24.24% of cases, no measurable impact is observed.** Regarding the XGBoost algorithm, we observe improvements in 60% of the cases across all models, except for the PF project, where MAE and SA significantly decrease by factors of 0.99 and 0.96, respectively. This improvement is most pronounced in the CP project, where SA performance increases up to 2.25 times, rising from 0.16 to 0.25. For ExtraTree and Random Forest, we observe significant improvements or no negative impact in 80% and 60% of cases, respectively, while in 20% and 40% of cases, the decrease remains statistically insignificant in 94% of the cases for both algorithms. For instance, in the MP project with random forest, MSE improves by a factor of 1.02, decreasing from 0.22 to 0.20.

**Kendall's tau results indicate weak similarity in 6.67% of cases and moderate similarity in 40% of cases, highlighting the substantial impact of removing deviations on feature importance rankings.** As shown in Table VI, removing deviations impacts the interpretability of XGBoost in three out of five projects, with the CP project experiencing the most substantial dissimilarity, dropping to 0.16. Similarly, for ExtraTree, we observe a moderate similarity in four out of five projects. In contrast, the feature importance rankings in Random Forest are less impacted, as all Kendall's tau values are strong, indicating high similarity. However, since none of these values equal one, some features still experience minor shifts in their importance.

**Regarding Top-k features commonly used in the literature, we observe weak overlap in 26.67% of cases and small overlap in 33.33% of cases, underscoring the necessity of considering deviation removal when analyzing the code review process.** More specifically, **for Top-1** features, we observe that the most important variable changes for all algorithms in the CP project. For instance, in the CP project with ExtraTree, the most critical variable shifts from the number of files to the number of commits, which alters its interpretability. The number of files could indicate the extent of modification, whereas the number of commits reflects the frequency of contributions. This change can impact how review complexity is assessed.

For **Top-3** features, we observe negligible overlap in 46.67% of cases and small overlap in 33.33%, indicating a substantial impact. A concrete example, in the DP project with XGBoost, the top three features shift from *title length, description length, associated sprint* to *number of commits, associated sprint, number of historical open MRs*, demonstrating that deviation removal significantly influences feature selection. Similarly, **for Top-5** features, we observe negligible overlap in 33.33% of cases and small overlap in 66.67%, further emphasizing the extensive impact across all case studies. These results reinforce the importance of considering deviations when interpreting the code review process.

## VI. DISCUSSION

**Generalizability:** *(a) Industry:* To validate our findings and ensure the robustness of our approach, we solicited data from another industrial partner. This company, which provides

TABLE V
METRICS FOR CODE REVIEW COMPLETION TIME PREDICTION

| Metric | Description | Metric | Description |
|---|---|---|---|
| Delta time | Time between project creation and MR creation | Associated sprint | the sprint Id in which the MR was created |
| #Committers | Number of MR committers | Has_Assignee | Indicates if MR has an assignee reviewer |
| #Authored_Commits | Number of commits authored before the MR creation | #Committed_commits | Number of changes committed |
| Change churn | MR total lines of code added+removed | Additions | Number of lines added |
| Deletions | Number of lines deleted | Title length | Length of MR title |
| Description length | Length of MR description | Is_Hashtag | Indicates if MR description contains a hashtag |
| Is_AtTag | Indicates if MR description contains an @mention | Source branch MRs | Number of open MRs on the same source branch |
| Source branch approval Time | Avg. approval time for source branch MRs | Target branch MRs | Number of open MRs targetting same branch |
| Target branch approval time | Avg. approval time for target branch MRs | | |
| #Minor contributors | Number of MR contributors with less than 5% of prior changes | #Major contributors | Contributors with more than 5% of prior changes |
| Self approved MRs | Number of historically self-approved MRs | MRs without_discussion | Number of historical MRs with no discussion |
| Avg. historical approval time | Avg. approval time for historical MRs | #Open MRs_history | Number of historical open MRs |
| Historical MR Size | Avg. size of past MRs | Historical entropy | Measure of historical change diversity |
| #Initial files | Number of files | | |

TABLE VI
PERFORMANCE AND FEATURE IMPORTANCE METRICS FOR DIFFERENT MODELS

| Model | XGBoost | | | | | | | ExtraTree | | | | | | | Random Forest | | | | | | |
|---|---|---|---|---|---|---|---|---|---|---|---|---|---|---|---|---|---|---|---|---|---|
| | Performance | | | Feature Importance | | | | Performance | | | Feature Importance | | | | Performance | | | Feature Importance | | | |
| | MSE | MAE | SA | K | T1 | T3 | T5 | MSE | MAE | SA | K | T1 | T3 | T5 | MSE | MAE | SA | K | T1 | T3 | T5 |
| PF | 0.99 | *0.99+* | *0.96++* | *0.74++* | 1++ | 1++ | 0.42* | **1.02** | **1** | 1.01 | 0.78++ | 1++ | 1++ | 0.42* | 0.99 | 0.99 | 1 | 0.81++ | 1++ | 1++ | 0.42* |
| DP | **1.02** | **1.01** | **1++** | 0.35+ | 1++ | 0.2 | 0.25 | 1 | 1 | *0.98++* | 0.49+ | 1++ | 0.2 | 0.25 | *1+* | **1** | **1** | 0.66++ | 1++ | 0.5* | 0.42* |
| FPGA | 1.01 | 1.01 | **0.98** | 0.67++ | 1++ | 0.50+ | 0.42* | **1.01** | **1** | 0.92 | 0.54+ | 1++ | 0.50* | 0.42* | 0.99 | 0.99 | 0.84 | 0.67++ | 1++ | 0.50* | 0.42* |
| MP | **1.02** | **1.03** | **1.07** | 0.60+ | 1++ | 0.50+ | 0.25 | *1.01++* | 0.99 | 0.99 | 0.55+ | 1++ | 0.20 | 0.11 | 1.02 | 1.00 | *0.99++* | 0.71++ | 1++ | 0.20 | 0.42* |
| CP | **1.10** | **1.03** | **2.25** | 0.16* | 1++ | 0.20 | 0.11 | **1.08** | **1.02** | 1 | 0.50+ | 0 | 0.20 | 0.42* | **1.11** | **1.02** | **1.09** | 0.66++ | 0 | 0.20 | 0.42* |

**Cohen's d**: Negligible if $|d| \leq 0.147$, Small+ if $0.147 < |d| \leq 0.33$, Medium++ if $0.33 < |d| \leq 0.474$, **Large** if $|d| > 0.474$.
**Wilcoxon Test**: *Significant* if $p - value \leq 0.05$
**Kendall's Tau (k)**: Weak* if $|\tau| \leq 0.3$, Moderate+ if $0.3 < |\tau| \leq 0.6$, Strong++ if $|\tau| > 0.6$.
**Top-K (Tk: T1, T2, T3) Overlap**: Negligible if $0.0 < Tk \leq 0.25$, Small* if $0.25 < Tk \leq 0.50$, Medium+ if $0.50 < Tk \leq 0.75$, Large++ if $0.75 < Tk \leq 1$.

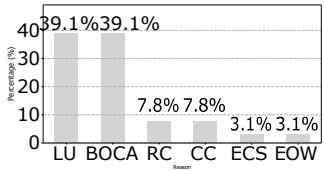

Fig. 4. Percentage distribution of deviation types in the project from the other industrial partner. *Note: BOCA = Build or Configuration Adjustments, EOW = Experimental or Work in Progress, CC = Code Cleaning, LU = Library Update, RC = Revert Commits, HC = Huge Changes, ECS = Empty Change Set.*

a Software-Defined Network (SDN) solution, allows us to extend our analysis beyond our initial dataset. Following the same process of MR annotation, we collaborate with their industry experts to identify deviations in one of their most critical projects, which contains 9k MRs.

***Our analysis revealed that the same deviations occurred in 17.88% of cases.*** As shown in Figure 4, we observe a similar pattern to our previous findings, where Library Updates (LU) accounted for 39.1% of the deviations, followed by Build or Configuration Adjustments (BOCA) at 39.1%. This high proportion can be attributed to the nature of the company's development process. The industry expert played a crucial role in distinguishing between normal cases and those that did not require review during dataset annotation. These findings emphasize the importance for practitioners to analyze similar patterns in their own datasets, as they may influence review analysis and related decisions.

***(b) Open Source:*** Analyzing deviations in open-source projects is challenging due to the lack of direct industrial partners for validation. In industry, deviations can be con- firmed through collaboration with practitioners who provide context on project workflows and MR usage. In contrast, open-source projects follow a decentralized model with di- verse contributors, making validation dependent on metadata, commit messages, and review patterns. Without direct access to maintainers, distinguishing true deviations from standard variations becomes more difficult.

Despite these challenges, applying our methodology to an open-source project is valuable, as it allows us to assess the generalizability of deviation detection beyond controlled industrial environments. By replicating our approach, we ana- lyze the Inkscape project, which includes 6,989 MRs, making it a suitable case study due to its active community-driven development and diverse MR usage. As shown in Figure 5, our analysis shows that deviations account for 10.69% of cases, though not all deviation types are equally prominent. Notably, experimental or WIP MRs represent the largest share (43.2%) [3] [4], followed by empty change set (29.7%) [5] and code cleaning (24.3%)[6]. Revert commits, on the other hand, are relatively rare, comprising only 2.7% of deviation cases [7]. Interestingly, an additional deviation type emerges in this con- text: **Documentation Updates**, which include MRs related to documentation modifications and language translations where

---
[3]https://gitlab.com/inkscape/inkscape/-/merge_requests/2467
[4]https://gitlab.com/inkscape/inkscape/-/merge_requests/747
[5]https://gitlab.com/inkscape/inkscape/-/merge_requests/457
[6]https://gitlab.com/inkscape/inkscape/-/merge_requests/3475
[7]https://gitlab.com/inkscape/inkscape/-/merge_requests/3342

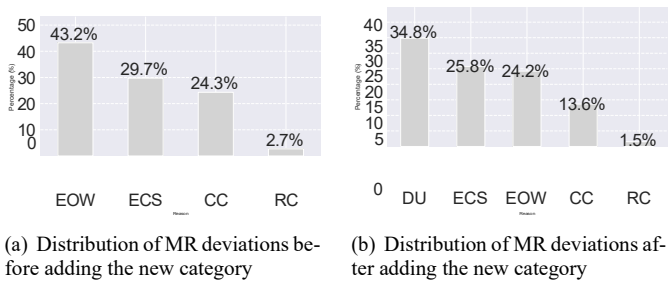

(a) Distribution of MR deviations before adding the new category

(b) Distribution of MR deviations after adding the new category

Fig. 5. Percentage Distribution of deviation Types for Inkscape
Note: BOCA = Build or Configuration Adjustments, EOW = Experimental or Work in Progress, CC = Code Cleaning, LU = Library Update, RC = Revert Commits, HC = Huge Changes, ECS = Empty Change Set, DU = Documentation Updates.

we do not observe any review activity[8] [9]. With this new deviation, MR deviations account for 16% of total deviations. ***This finding highlights the importance of investigating deviation patterns within different contexts, encouraging practitioners to examine deviations specific to their own projects.*** Notably, The overall percentage of deviations in Inkscape is lower than in the industrial dataset, which we attribute to differences in workflow structures, review policies, and project governance. In open-source projects, we hypothesize that review deviations are naturally less frequent due to their more flexible and asynchronous review processes, whereas in industrial settings, structured workflows and predefined roles contribute to a higher incidence of deviations.

**Explaining MR deviations:** The MR deviations identified in RQ1 do not conform to the standard code review process, with each type of deviation exhibiting distinct characteristics and behaviors. These deviations account for up to 37.02% of the total MRs, potentially distorting key insights into review dynamics. Understanding these deviations individually provides critical supplementary information that would otherwise remain hidden. Ignoring these MRs would exclude a substantial portion of project activity, which is why we conducted additional analyses to investigate their interpretation compared to standard MRs.

From a statistical perspective, an analysis of code review completion time did not reveal a significant difference between standard and deviation MRs. However, ***our machine learning experiments with 100 bootstraps uncover notable distinctions in feature importance***. Specifically, when comparing models trained on deviation MRs versus regular MRs, XGBoost (the best-performing model in RQ3, with results consistent across other algorithms) exhibits distinct behaviors. The Kendall's tau correlation between models trained on standard and deviation MRs ranges from 0.26 to 0.64, indicating varying degrees of divergence. Additionally, while Top-1 feature rankings remain stable across all projects, Top-3 feature overlap is 0.5 in four projects and 0.2 in one project. Finally, Top-5 feature overlap is 0.25 in three projects and 0.42 in two projects, highlighting the unique patterns embedded in MR deviations compared to regular MRs. The observed differences suggest that treating deviation MRs separately in predictive models could enhance code review analysis by introducing an additional analytical dimension, enabling deeper insights from machine learning models.

**Model Validation:** To assess the robustness and generalizability of our RQ2 models, we conducted cross-validation across different projects, training on one project (e.g., CP) and validating on another (e.g., DP). This approach evaluates the model's ability to adapt to unseen data and ensures that it is not overfitting to a specific project's characteristics. Our results indicate accuracy ranging from 0.54 to 0.72, highlighting the model's adaptability while also revealing variability in performance across different contexts.

## VII. CONCLUSION

This study addresses a critical gap in code review research by identifying and categorizing deviations—MRs that do not follow the expected review workflow—in industrial software development. We show that deviations account for up to 37.02% of MRs and can distort analytical conclusions if not properly handled. To address this, we develop an AI-driven approach using few-shot learning to automate deviation detection, achieving up to 91% accuracy, significantly reducing manual effort while maintaining high precision. Moreover, we demonstrate that excluding deviations improves or stabilizes ML model performance in predicting code review completion time in 77.57% of cases. Additionally, deviation removal impacts overall feature importance in 47% of cases and top-k feature rankings in 60% of cases, highlighting its influence on model behavior. These results confirm that deviation elimination mitigates biases that could otherwise lead to misleading process optimizations. Our findings underscore the necessity of contextual analysis in MR evaluations to ensure that review efforts focus on meaningful code changes. By integrating deviation detection into the review pipeline, teams can enhance efficiency, prioritize critical MRs, and improve the reliability of code review analytics.

**Threats to Validity:** Our study focuses on MR mechanism. While other code review mechanisms exist, they generally follow the same fundamental principles of peer review, discussion, and approval. Therefore, we believe that our findings are not significantly impacted by the choice of the mechanism but rather by the context of the project, which may introduce new types of deviations specific to its development practices and team dynamics. Another potential threat lies in the technology stack and tools used for our analysis. To mitigate this risk, we ensured the use of well-established and reliable tools and libraries, such as scikit-learn. This choice enhances the reproducibility and trustworthiness of our results.

**Acknowledgement:** We extend our sincere gratitude to **Seyedbehnam Mashari** for their invaluable contribution and collaboration. Their expertise in MR analysis greatly enhanced the reliability of our annotation phase. We also acknowledge **Bassem Guendy** for their continuous support and guidance throughout this study. Their expertise was instrumental in providing critical insights in our work, facilitating data access, and assisting in the validation of our results, ensuring the robustness of our analysis.

---

[8] https://gitlab.com/inkscape/inkscape/-/merge_requests/2295
[9] https://gitlab.com/inkscape/inkscape/-/merge_requests/33264